\documentclass[english]{article}
\usepackage[T1]{fontenc}
\usepackage[latin9]{inputenc}
\usepackage{amsbsy}
\usepackage{esint}

\makeatletter
\usepackage{fullpage}

\makeatother

\usepackage{babel}
\begin{document}

\title{Weyl gravity as general relativity}

\author{James T. Wheeler%
\thanks{Utah State University Department of Physics 4415 Old Main Hill, Logan,
UT 84322-4415; jim.wheeler@usu.edu%
}}
\maketitle
\begin{abstract}
When the full connection of Weyl conformal gravity is varied instead
of just the metric, the resulting vacuum field equations reduce to
the vacuum Einstein equation, up to the choice of local units, if
and only if the torsion vanishes. This result differs strongly from
the usual fourth-order formulation of Weyl gravity. 
\end{abstract}

\section*{Introduction}

General relativity accounts in exquisite detail for nearly all gravitational
phenomena. Thorough tests of its predictions have met with repeated
success, as alternative theories have required modification or abandonment.
Still, some alternative theories survive and provide, at the very
least, further tests of our understanding. From the introduction of
the equivalence principle in 1908, through the presentation of the
final field equations in 1915, development of general relativity was
rapid \cite{MTW}, yet no faster than the introduction of a second
relativistic theory of gravity by Weyl, Bach and others in the years
from 1918 to 1924. This second theory, called conformal gravity or
Weyl gravity, remains a topic of active discussion despite its higher
order field equations. We hope to clarify the reason why it has been
difficult to distinguish these first two relativistic gravity theories.
Specifically, we show that the when all of the connection fields of
conformal gravity are varied independently instead of the usual fourth-order,
metric-only variation, the torsion-free solutions of the two theories
differ only in that the field equations of conformal gravity are unchanged
by the use of arbitrary local choices of units. 

Beginning in 1918, Weyl \cite{Weyl 1918a}, \cite{Weyl 1918b} and
Bach \cite{Bach 1921} developed a theory of gravity based on the
conformally invariant Weyl action

\begin{equation}
S=\alpha\int C_{abcd}C^{abcd}\sqrt{-g}d^{4}x\label{Action}
\end{equation}
Bach adds to this a second term quadratic in the dilatational curvature.
Variation leads to the Bach equation,
\begin{equation}
D_{\alpha}D_{\beta}C^{\mu\alpha\nu\beta}-\frac{1}{2}C^{\mu\alpha\nu\beta}R_{\alpha\beta}=0\label{Bach equation}
\end{equation}
An alternate approach \cite{Bach 1921} makes use of the Gauss-Bonnet
integral for the Euler character \cite{Frenchel 1940} to write the
equivalent action, $S'=2\alpha\int\left(R_{ab}R^{ab}-\frac{1}{3}R^{2}\right)\sqrt{-g}d^{4}x$,
and field equations depending only on the Ricci tensor and its derivatives.

From the point of view of quantum theory, Weyl conformal gravity has
an important advantage and an equally important disadvantage. Power
counting suggests that the curvature-quadratic action, eq.(\ref{Action}),
is renormalizable. However, the presence of fourth order derivatives
in the field equations, eq.(\ref{Bach equation}), is generally associated
with non-unitarity. Rather than entering into the controversy surrounding
these observations (see, e.g., \cite{Bender Mannheim 2008}, \cite{Smigla 2009}),
we propose a full connection variation of eq.(\ref{Action}). We show
that torsion-free solutions to the resulting field equations lead
purely to the second order field equation of general relativity, modified
to have local dilatational covariance. Within this alternative approach,
the debate over unitarity becomes moot.

Discussion also surrounds certain solutions to the Bach equation.
Bach's generalization of the Schwarzschild solution \cite{Bach 1921},
for example, has been developed into a model to explain galactic rotation
curves \cite{Mannheim 2006}, but may fail at solar system scales
\cite{Flanagan 2006}, \cite{Sultana et al 2012}. This discussion
has faded in importance as many more independent consequences and
tests of dark matter have emerged. Again, within our current presentation,
these considerations do not apply.

It is interesting to speculate that conformal gravity with full connection
variation, having a dimensionless action, might give rise to a renormalizable
quantization of general relativity or contribute to a deeper understanding
of the relevance of twistor string theory \cite{Isberg 1992}.

Solutions for the metric in conformal gravity are determined only
up to an overall multiple, forming elements of conformal equivalence
classes, $g_{\alpha\beta}\in\left\{ e^{2\varphi}g_{\alpha\beta}\mid all\,\varphi\left(x^{\mu}\right)\right\} $.
As long as the dilatational potential, the Weyl vector, is a pure
gradient, it is consistent to regard this factor as a choice of local
units. Given this requirement for conformal equivalence classes of
solutions, it becomes necessary to ask when a given metric is conformal
to a metric satisfying the Einstein equation. This question was first
addressed in 1924 by Brinkmann \cite{Brinkmann 1924}, who found a
set of necessary and sufficient conditions for a space to satisfy
the vacuum Einstein equation up to a conformal transformation. These
expressions have the disadvantage of depending on the conformal transformation
itself, so that one simultaneously checks for the existence of a suitable
transformation and finds it. In 1963, Szekeres \cite{Szekeres 1963}
used spinor techniques to separate the \emph{existence} of a conformal
transformation to an Einstein space from the problem of \emph{finding}
that transformation. As expected from the fact that a conformal transformation
changes the Ricci tensor by terms involving the second derivative
of the conformal factor, there are two integrability conditions. Subsequent
work refines or gives different expression to these results \cite{Kozameh et al 1985}.

It is crucial to the present investigation that the Bach equation
(for which torsion is \emph{always} assumed to vanish) has solutions
which are \emph{not} solutions to the vacuum Einstein equation. The
need for equivalence classes of metrics complicates this. It was not
until 1984 that Schmidt \cite{Schmidt 1984} showed conclusively the
existence of solutions to conformal gravity which are not conformally
equivalent to Einstein spaces (i.e., spaces for which the Ricci tensor
equals a constant times the metric). Subsequently, additional non-Einstein
space solutions were found by Nurowski and Plebanski \cite{Nurowski 2001},
and six more solutions by Liu, Lü, Pope and Vázquez-Poritz \cite{Liu et al 2013}.
The existence of non-Einstein solutions to fourth-order conformal
gravity demonstrates that the stronger restrictions that we describe
here are not vacuous \textendash{} our results below demonstrate a
distinct interpretation of Weyl gravity from the fourth-order theory.
Since our method is natural within the context of conformal gauge
theory, we will refer to conventional conformal gravity as the fourth-order
theory and the method we employ as \emph{auxiliary conformal gauge
theory}. The name stems from the way the special conformal gauge fields
act as auxiliary fields that turn the full curvature into the Weyl
curvature.

Conformal gauge theory was first written down in the mid-1970s. Leading
up to a conformal supergravity model, Crispim-Romao, Ferber and Freund
\cite{Crispim Romao et al 1977} performed the first gauging of the
conformal group, $O\left(4,2\right)$, writing the Weyl action in
terms of the conformal curvatures. Kaku, Townsend and van Nieuwenhuizen
\cite{Kaku et al 1977} developed a similar gauging. Both this group
\cite{Kaku et al 1978}, and Crispim-Romao \cite{Crispim-Romao 1978}
went on to write superconformal gravity theories (for a review of
superconformal gravity, see \cite{Fradkin Tseytlin}), and both show
that the gauge field of the special conformal transformations is the
Schouten tensor (equivalent to the Ricci tensor) hence auxiliary (see
also \cite{Wheeler PRD auxiliary}). Within a few years, Ivanov and
Niederle \cite{Ivanov Niederle I 1982,Ivanov Niederle II 1982} gave
a more systematic treatment of gauge theories of gravity, using techniques
developed by Ne'eman and Regge \cite{Neeman and Regge} based on the
work of Cartan \cite{Kobayashi and Nomizu}. Their work identifies
two distinct conformal gaugings, now called the \emph{auxiliary} and
\emph{biconformal} \cite{Wheeler JMP},\cite{WW} gaugings.

We use these techniques in our formulation since they have the advantage
of treating each independent gauge field on an equal footing. This
makes variation of all 15 gauge vectors natural, giving additional
field equations beyond the Bach equation. Half the equations are easily
solved, establishing the auxiliary field and showing the equivalence
to Weyl gravity. These results are well-known. However, the variation
of the spin connection provides another field equation, the vanishing
divergence of the Weyl curvature. \emph{Our central result is to show
that this equation is an integrability condition which reduces the
theory to scale-invariant general relativity}. With this change from
varying only the metric to varying all of the connection fields, Weyl
gravity changes from a fourth order theory into a theory of conformal
equivalence classes of solutions to ordinary general relativity.

In the next Section, we develop auxiliary conformal gravity. Though
our action is not initially invariant under the full conformal group,
it is well-known that the field equation of the special conformal
transformation gauge field reduces the action to eq.(\ref{Action}).
From the scale-invariant action, we could perform either the fourth-order
metric variation by assuming the metric form of the connection, or
the gauge theory approach in which each connection component is independently
varied. Writing the field equations from the latter, we show that
any torsion-free solution of the new field equations solves the Bach
equation. Finally, we show that the new field equation is the integrability
condition that forces solutions to be conformal equivalence classes
of solutions to the vacuum Einstein equation.

\section*{Auxiliary conformal gauge theory}

We briefly outline auxiliary conformal gauge theory, culminating in
the action and field equations. The basic construction uses group
quotients to construct a fiber bundle with chosen symmetry, then modifies
the base manifold and connection to give curvature (\cite{Ivanov Niederle I 1982}-\cite{Kobayashi and Nomizu}).
The advantage of the approach is that it keeps the curvatures and
action expressed in terms of the gauge fields, making the variation
straightforward. In the next Section we consider solutions.

The conformal group of spacetime has fifteen generators: six for Lorentz
transformations, four translations, four special conformal transformations,
and one dilatation. For each of these we have a corresponding dual
$1$-form: $\boldsymbol{\omega}^{A}\in\left\{ \boldsymbol{\omega}_{\; b}^{a},\mathbf{e}^{a},\mathbf{f}_{a},\boldsymbol{\omega}\right\} $
called the spin connection, the solder form, the gauge field of special
conformal transformations and the Weyl vector, respectively. These,
together with the group structure constants, are substituted into
the Maurer-Cartan equation.

To recover Weyl gravity, we take the quotient of the conformal group
by the inhomogeneous Weyl subgroup, $\mathcal{IW}$, generated by
Lorentz transformations, special conformal transformations, and dilatations.
This quotient is a homogeneous, 4-dim manifold, and the 1-forms above
provide its connection. Next, we modify this structure by generalizing
the manifold, and by changing the connection. Changing the manifold
has no effect on the local structure, but changing the connection
modifies the Maurer-Cartan equation, resulting in the addition of
curvature $2$-forms, $\boldsymbol{\Omega}^{A}\in\left\{ \boldsymbol{\Omega}_{\; b}^{a},\mathbf{T}^{a},\mathbf{S}_{a},\boldsymbol{\Omega}\right\} $.
We place two restrictions on these curvatures. First, we require the
curvatures to characterize the manifold only. In general, an integral
of the connection along a curve in the full space gives a conformal
transformation, with integrals around closed loops equivalent to surface
integrals of the curvatures. We require horizontlity: these closed
loop integrals must be independent of the $\mathcal{IW}$ subgroup
transformations, which occurs if and only if the curvatures may be
expanded in terms of the solder forms, $\boldsymbol{\Omega}^{A}=\frac{1}{2}\Omega_{\; cd}^{A}\mathbf{e}^{c}\wedge\mathbf{e}^{d}$,
and not all fifteen connection forms. Second, we require integrability
of the Cartan equations (i.e., these modified Maurer-Cartan equations).
This leads to Bianchi identities for the curvatures. The Cartan equations
and Bianchi identities are given in Appendix A.

The quotient construction describes only the geometry, leading us
to the form for the curvatures in terms of the gauge fields, which
agree with those found in \cite{Kaku et al 1978}, \cite{Crispim-Romao 1978},
and \cite{Ivanov Niederle I 1982}, and proivding the Bianchi identities.
The physical content arises solely from the field equations, found
by writing an action functional defined on the local $\mathcal{IW}$-invariant
principal bundle. The action is constructible from the available tensors,
$\mathbf{e}^{c},\boldsymbol{\Omega}_{\; B}^{A}$, together with the
invariant metric and Levi-Civita tensors, $\eta_{ab},\varepsilon_{abcd}$.
Scale invariance requires curvature-quadratic terms, and the most
general even parity, $\mathcal{IW}$-invariant possibility is uniquely
determined (up to an overall multiple) to be
\begin{eqnarray*}
S_{auxiliary}^{\mathcal{IW}} & = & \alpha\int\boldsymbol{\Omega}_{\; B}^{A}\wedge{}^{*}\boldsymbol{\Omega}_{\; A}^{B}\\
 & = & \alpha\int\left(\boldsymbol{\Omega}_{\; b}^{a}\wedge{}^{*}\boldsymbol{\Omega}_{\; a}^{b}+4\mathbf{T}^{c}\wedge{}^{*}\mathbf{S}_{c}+2\boldsymbol{\Omega}\wedge{}^{*}\boldsymbol{\Omega}\right)
\end{eqnarray*}
where $\boldsymbol{\Omega}_{\; B}^{A}$ is the full $SO\left(4,2\right)$
curvature $2$-form. This does not lead to Weyl gravity, as will be
shown elsewhere, but instead to a Weyl-Cartan geometry (i.e., one
having nontrivial dilatation and torsion). To achieve Weyl gravity
on the $\mathcal{IW}$ bundle, we need to break the special conformal
symmetry with our choice of the action. Since the curvature has already
broken the translational symmetry, we expect both non-dynamical torsion
and non-dynamical special conformal curvature. Dropping the center
term in $S_{auxiliary}^{\mathcal{IW}}$, we have the $\mathcal{W}$-invariant
Weyl-Bach action,
\begin{eqnarray}
S_{auxiliary}^{\mathcal{W}} & = & \int\left(\alpha\boldsymbol{\Omega}_{\; b}^{a}\wedge{}^{*}\boldsymbol{\Omega}_{\; a}^{b}+\beta\boldsymbol{\Omega}\wedge{}^{*}\boldsymbol{\Omega}\right)\label{eq: W invariant action}
\end{eqnarray}
The equivalence between the first term and the original conformal
gravity action is established in \cite{Kaku et al 1978} and \cite{Ivanov Niederle I 1982},
while the vanishing of the second term is shown below. Bach's original
action included both terms but with the critical value of $\beta=2\alpha$
(in our notation). A detailed discussion of these symmetries is provided
in Appendix B.

Varying the entire Cartan connection gives the field equations. This
is where the difference between our approach and the usual approach
to Weyl gravity occurs. In order to display the torsion dependence
of the field equations explicitly, we write the field equations in
a coordinate basis.
\begin{eqnarray}
D_{\tau}\Omega_{\;\nu}^{\mu\;\lambda\tau}+\Omega_{\;\nu}^{\mu\;\lambda\alpha}T_{\;\alpha\tau}^{\tau}+\frac{1}{2}\Omega_{\;\nu}^{\mu\;\alpha\tau}T_{\;\alpha\tau}^{\lambda} & = & 0\label{eq: Lorentz field equation}\\
D_{\nu}\Omega^{\mu\nu}+\Omega^{\mu\alpha}T_{\;\alpha\nu}^{\nu}+\frac{1}{2}\Omega^{\alpha\nu}T_{\;\alpha\nu}^{\mu} & = & 0\label{eq: Dilatational field equation}\\
2\alpha\Omega_{\;\;\alpha\mu\beta}^{\mu}+\beta\Omega_{\alpha\beta} & = & 0\label{eq: Special conformal field equation}\\
2\alpha f_{\mu\nu}\Omega^{\mu\alpha\nu\beta}+\beta f_{\;\mu}^{\alpha}\Omega^{\beta\mu} & = & -\alpha\Theta^{\alpha\beta}-\beta Q^{\alpha\beta}\label{eq:Solder form field equation}
\end{eqnarray}
where eq.(\ref{eq: Lorentz field equation}) arises from the variation
of $\boldsymbol{\omega}_{\; b}^{a}$, eq.(\ref{eq: Dilatational field equation})
from $\boldsymbol{\omega}$, eq.(\ref{eq: Special conformal field equation})
from $\mathbf{f}_{a}$ and eq.(\ref{eq:Solder form field equation})
from $\mathbf{e}^{a}$. All occurences of the torsion arise from derivatives
of the solder form. The sources for the solder form equation are given
by,
\begin{eqnarray}
\Theta^{\alpha\beta} & \equiv & -\Omega^{\mu\nu\rho\alpha}\Omega_{\mu\nu\rho}^{\;\quad\beta}+\frac{1}{4}\Omega^{\mu\nu\rho\sigma}\Omega_{\mu\nu\rho\sigma}g^{\alpha\beta}\label{Energy momentum of Weyl curvature}\\
Q^{\alpha\beta} & \equiv & \Omega^{\mu\alpha}\Omega_{\mu}^{\;\beta}-\frac{1}{4}\Omega^{\mu\nu}\Omega_{\mu\nu}g^{\alpha\beta}\label{Energy momentum of dilatation}
\end{eqnarray}
These sources arise because the Hodge dual is a nonlinear function
of the solder form. The covariant derivatives are taken using the
torsionful, Weyl-covariant, metric compatible connection
\[
\tilde{\Gamma}_{\;\mu\nu}^{\beta}\equiv\Gamma_{\;\mu\nu}^{\beta}-\left(\delta_{\mu}^{\beta}W_{\nu}+\delta_{\nu}^{\beta}W_{\mu}-g^{\alpha\beta}g_{\nu\mu}W_{\alpha}\right)+\frac{1}{2}\left(T_{\mu\;\nu}^{\;\beta}+T_{\nu\;\mu}^{\;\beta}-T_{\;\mu\nu}^{\beta}\right)
\]
where $\Gamma_{\;\mu\nu}^{\beta}$ is the usual Christoffel connection.
The derivation of this and other useful relations is described in
Appendix C.

Having expressed the field equations in terms of covariant derivatives
satisfying $D_{\alpha}e_{\beta}^{\quad a}=0$, we may freely interchange
between coordinate (Greek) and orthonormal (Latin) indices.

\section*{Solving the field equations}

The system to be solved now consists of eqs.(\ref{eq: Lorentz field equation}-\ref{Energy momentum of dilatation})
with the form of the curvatures dictated by the conformal group. Our
central result is to show that the solution is scale-invariant general
relativity if and only if the torsion vanishes. Of course, if the
torsion does not vanish we do not have general relativity. To complete
the result, we must show that when we set the torsion to zero, $T_{\;\alpha\beta}^{\mu}=0$,
eqs.(\ref{eq: Lorentz field equation}-\ref{eq:Solder form field equation})
describe dilatationally covariant general relativity. 

First, we show that the dilatational curvature, $\Omega_{ab}$, generically
vanishes. We simplify the special conformal field equation, eq.(\ref{eq: Special conformal field equation})
using the torsion-free Bianchi identity (see Appendix A) of the solder
form, $\Omega_{\;\left[bcd\right]}^{a}=\delta_{[b}^{a}\Omega_{cd]}$.
Combining its trace, $\Omega_{\; bac}^{a}-\Omega_{\; cab}^{a}=-2\Omega_{bc}$,
with the antisymmetric part of field equation, eq.(\ref{eq: Special conformal field equation}),
leads to 
\[
\left(2\alpha-\beta\right)\Omega_{ab}=0
\]
Unless the arbitrary constants in the action are related by $\beta=2\alpha$,
the dilatation vanishes $\Omega_{ab}=0$ and there is no physically
measurable size change. Eq.(\ref{eq: Dilatational field equation})
is now satisfied.

Next, we find the special conformal gauge field, $f_{ab}$. With vanishing
dilatational curvature, eq.(\ref{eq: Special conformal field equation})
reduces to $\Omega_{\; abc}^{b}=0$. Defining the Riemann curvature
of the spin connection 
\begin{eqnarray*}
\mathbf{R}_{\quad b}^{a} & \equiv & \mathbf{d}\boldsymbol{\omega}_{\; b}^{a}-\boldsymbol{\omega}_{\; b}^{c}\wedge\boldsymbol{\omega}_{\; c}^{a},
\end{eqnarray*}
the Lorentz curvature becomes $\Omega_{\; bcd}^{a}=R_{\; bcd}^{a}+\delta_{c}^{a}f_{bd}-\delta_{d}^{a}f_{bc}+f_{\; c}^{a}\eta_{bd}-f_{\; d}^{a}\eta_{bc}$.
Substituting into the field equation, we readily solve,
\begin{eqnarray*}
f_{ab} & = & -\frac{1}{2}\left(R_{bd}-\frac{1}{6}R\eta_{ab}\right)\equiv-\mathcal{R}_{ab}
\end{eqnarray*}
where the \emph{Schouten tensor} $\mathcal{R}_{ab}$ may be used interchangeably
with the Ricci tensor, since $R_{ab}=2\mathcal{R}_{ab}+\eta_{ab}\mathcal{R}$.
This result is well-known (\cite{Kaku et al 1978}, \cite{Crispim-Romao 1978},
\cite{Wheeler PRD auxiliary}) and it eliminates $f_{ab}$ from the
problem. Substituting into the Lorentz curvature yields the Weyl curvature,
$\Omega_{\; bcd}^{a}=C_{\; bcd}^{a}$, so the ``auxiliary'' field,
$f_{ab}$, systematically enforces the conformal structure. Also,
substituting $f_{ab}=-\mathcal{R}_{ab}=f_{\left(ab\right)}$ into
the expression for the dilatation in terms of the connection, $\Omega_{ab}=f_{\left[ab\right]}+\omega_{\left[a;b\right]}$
shows that the Weyl vector $\omega_{a}$ is a pure gradient.

These considerations show the equivalence of the auxiliary and Weyl
actions, so the Bach equation must be satisfied. To see how, first
observe that $\Theta_{ab}$, eq.(\ref{Energy momentum of Weyl curvature}),
becomes the energy momentum tensor of the Weyl curvature and therefore
vanishes identically in $4$-dim. This identity, $\Theta_{ab}=0$
was first shown by Lanczos \cite{Lanczos 1938} (but see also \cite{Lovelock 1970},
\cite{Trujillo-Wheeler}). Vanishing dilatation gives $Q^{\alpha\beta}=0$
in eq.(\ref{Energy momentum of dilatation}), reducing the right side
of eq.(\ref{eq:Solder form field equation}) to zero. Finally, replacing
$\Omega^{abcd}=C^{abcd}$ and $f_{ab}=-\mathcal{R}_{ab}$, in eq.(\ref{eq:Solder form field equation})
gives $R_{cd}C^{cadb}=0$, which, combined with the covariant divergence
of eq.(\ref{eq: Lorentz field equation}), reproduces the Bach equation.

We conclude that all solutions to torsion-free auxiliary conformal
gauge theory are also solutions to fourth-order field Weyl gravity.
The converse, however, is not true. We now show that the class of
solutions is equal to the set of conformal equivalence classes of
Ricci-flat spacetimes.

The calculation centers on the vanishing divergence of the Weyl curvature,
$D_{a}C^{abcd}=0$, the torsion-free version of eq.(\ref{eq: Lorentz field equation}).
Clearly, this field equation distinguishes the two approaches to Weyl
gravity. Without this, we would only have vanishing dilatation, a
gauge with vanishing Weyl vector, and $f_{\mu\nu}=-\mathcal{R}_{\mu\nu}$
turning the curvature into the Weyl curvature, all of which reduces
the problem to the usual Weyl curvature squared action together with
the metric variation that gives the Bach equation. The remarkable
thing is that the vanishing divergence is an integrability condition
that reduces the fourth order theory to a second order theory. To
see this, we choose the conformal gauge so that the Weyl vector vanishes.
This makes the geometry appear Riemannian. The field equations reduce
to eq.(\ref{eq: Lorentz field equation}) and, from eq.(\ref{eq:Solder form field equation}),
the condition $R_{ab}C^{acbd}=0$. This latter expresses the vanishing
energy-momentum of the Schouten tensor \cite{Trujillo-Wheeler}.

Expanding the second Bianchi identity, $R_{\; b\left[cd;e\right]}^{a}=0$
in terms of the Weyl and Schouten parts then taking a trace relates
the divergence of the Weyl curvature to the Schouten tensor. Using
the identity $\mathcal{R}_{;d}=\mathcal{R}_{\; d;a}^{a}$, from a
second trace, the Bianchi identity becomes 
\begin{eqnarray*}
C_{\; bcd;a}^{a}+\left(n-3\right)\left(\mathcal{R}_{bc;d}-\mathcal{R}_{bd;c}\right) & = & 0
\end{eqnarray*}
(in $n$-dim) so the field equation may be written as
\begin{eqnarray}
\mathcal{R}_{b\left[c;d\right]} & = & 0\label{Field equation in Riemannian gauge}
\end{eqnarray}
This is \emph{not} the well-known integrability condition,
\begin{equation}
\mathcal{R}_{bc;d}-\mathcal{R}_{bd;c}+\varphi_{a}C_{\; bcd}^{a}=0,\label{Integrability condition}
\end{equation}
for the existence of a gauge in which the vacuum Einstein equation
holds (\cite{Brinkmann 1924}, \cite{Szekeres 1963}). The problem
is that we are in the wrong basis to see the integrability condition.
\emph{Staying in the Riemannian gauge, we define a new basis},
\[
\tilde{\mathbf{e}}^{a}=e^{\chi}\mathbf{e}^{a}
\]
and require the same relations between $\tilde{\mathbf{e}}^{a},\tilde{\boldsymbol{\omega}}_{\; b}^{a}$
and $\tilde{\mathbf{R}}_{\; b}^{a}$ as hold between $\mathbf{e}^{a},\boldsymbol{\omega}_{\; b}^{a}$
and $\mathbf{R}_{\; b}^{a}$. The Bianchi identity remains the same,
$\tilde{C}_{\; bcd;a}^{a}+\left(n-3\right)\left(\tilde{\mathcal{R}}_{bc;d}-\tilde{\mathcal{R}}_{bd;c}\right)=0$
but the field equation differs. With the new connection given by
\[
\tilde{\boldsymbol{\omega}}_{\; b}^{a}=\boldsymbol{\omega}_{\; b}^{a}+2\Delta_{db}^{ac}\chi_{c}\mathbf{e}^{d},
\]
we find the well-known change in the Schouten tensor \cite{Levi-Civita},
with the Weyl curvature unchanged. The divergence of the Weyl curvature,
however, is related to the old by $D_{a}^{\left(\omega\right)}C_{\;\; bcd}^{a}=e^{2\chi}\left(\tilde{D}_{a}\tilde{C}_{\;\; bcd}^{a}-\left(n-3\right)\chi_{e}\tilde{C}_{\;\; bcd}^{e}\right)=0$.
Combining this with the Bianchi identity, the field equation is now
\begin{eqnarray}
\tilde{\mathcal{R}}_{bc;d}-\tilde{\mathcal{R}}_{bd;c}+\chi_{e}\tilde{C}_{\;\; bcd}^{e} & = & 0\label{Integrability}
\end{eqnarray}
and this \emph{is} the integrability condition. Therefore, there exists
a gauge, $\chi$, which takes $\tilde{\mathbf{e}}^{a}$ to a Ricci-flat
basis.

Some care is now required. The integrability condition, eq.(\ref{Integrability condition}),
tells us that $\tilde{\mathbf{e}}^{a}$ is conformal to a basis in
which the spacetime has vanishing Ricci tensor. Let this basis be
$\hat{\mathbf{e}}^{a}=e^{\xi}\tilde{\mathbf{e}}^{a}$ for some function
$\xi$. Then, since $\tilde{\mathbf{e}}^{a}=e^{\chi}\mathbf{e}^{a}$,
we have $\hat{\mathbf{e}}^{a}=e^{\xi+\chi}\mathbf{e}^{a}$. This means
that the original basis is conformal to a Ricci-flat basis, and therefore
the integrability condition must hold there as well,
\[
\mathcal{R}_{bc;d}-\mathcal{R}_{bd;c}+\zeta_{e}C_{\;\; bcd}^{e}=0
\]
where $\zeta=\xi+\chi$. However, since we know that $\mathcal{R}_{bc;d}-\mathcal{R}_{bd;c}=0$
by the field equation we must also have
\[
\zeta_{e}C_{\;\; bcd}^{e}=0
\]
in the original basis. For spacetimes other than Petrov types O or
N, this only happens if
\[
\zeta_{a}\equiv e_{a}^{\quad\mu}\partial_{\mu}\zeta=0
\]
so $\zeta=\zeta_{0}$ is at most a constant and $\hat{\mathbf{e}}^{a}=e^{\zeta_{0}}\mathbf{e}^{a}$.
A constant multiplying the basis preserves the vanishing Ricci tensor,
so the Ricci tensor vanishes in the original basis. While Petrov type
O and N spaces are conformally Ricci flat, the Ricci tensor may not
vanish in Riemannian gauge. It is worth noting that a number of the
non-Einstein solutions in \cite{Liu et al 2013} are wavelike solutions
of type N. These special cases warrant further study; type $N$ spaces
are the same ones Szekeres found to be exceptional \cite{Szekeres 1963}.

We see that the Riemannian gauge is doubly special (except possibly
in type O or N spaces): \emph{both} the Weyl vector \emph{and} the
Ricci tensor vanish in that gauge. This is part of the reason the
integrability was not seen earlier. There is another reason as well,
stemming from the fact that the special form $\mathcal{R}_{bc;d}-\mathcal{R}_{bd;c}=0$
admits Einstein spaces as solutions. We now clarify this issue.

Eq.(\ref{Field equation in Riemannian gauge}) is solved by any Einstein
space, $\mathcal{R}_{ab}=\frac{1}{6}\Lambda\eta_{ab}$ for constant
$\Lambda$. This seems to contradict our result above. The problem
is resolved if we recall that solutions are conformal equivalence
classes of metrics, $\left\{ e^{2\varphi}g_{\alpha\beta}\mid all\;\varphi\right\} $.
Above, we showed that metrics conformal to Ricci-flat metrics comprise
such a class. If we compute the condition for a space to be conformal
to an Einstein space, however, eq.(\ref{Integrability condition})
gains a new term,

\[
\mathcal{R}_{a\left[b;c\right]}+\chi_{d}C_{\; abc}^{d}+\frac{1}{3}\Lambda\eta_{a[b}\chi_{c]}=0
\]
Since the field equation requires $\mathcal{R}_{a\left[b;c\right]}+\chi_{d}C_{\; abc}^{d}=0$,
a conformal equivalence class with cosmological constant also requires
\[
\frac{1}{3}\Lambda\eta_{a[b}\chi_{c]}=0
\]
Since $\chi$ is arbitrary, the cosmological constant must vanish,
$\Lambda=0$. We conclude that, while all Einstein spaces satisfy
eq.(\ref{Field equation in Riemannian gauge}), the only conformal
equivalence class of Einstein spaces satisfying eq.(\ref{Field equation in Riemannian gauge})
when expressed in the Riemannian gauge is the class with $\Lambda=0$,
and therefore the Ricci-flat equivalence class.

Finally, we return to the field equations in an arbitrary gauge, so
the Weyl vector is no longer zero. The Schouten tensor becomes
\begin{equation}
\mathcal{R}_{ab}=\mathcal{R}_{ab}^{\left(\alpha\right)}-\omega_{\left(a;b\right)}-\omega_{a}\omega_{b}+\frac{1}{2}\omega^{2}\eta_{ab}\label{Schouten Weyl tensor}
\end{equation}
where $\mathcal{R}_{ab}^{\left(\alpha\right)}$ is the Schouten tensor
in Riemannian gauge. But $\mathcal{R}_{ab}$ is covariant under conformal
transformations, with $\mathcal{R}_{ab}\rightarrow e^{-2\chi}\mathcal{R}_{ab}$.
We may therefore evaluate it in any convenient gauge and immediately
know it in any other. Now we see the importance of generically having
both vanishing Weyl vector and vanishing Ricci tensor in the Riemannian
gauge \textendash{} evaluating eq.(\ref{Schouten Weyl tensor}) in
the Riemannian gauge now shows that
\[
\mathcal{R}_{ab}=0
\]
in \emph{every} gauge.

In conclusion, we have shown that when all connection fields of conformal
gravity are varied independently, solutions are conformal equivalence
classes of solutions to the vacuum Einstein equation. Quantization
of conformal gravity may therefore be renormalizable, ghost free,
and essentially equivalent to general relativity. Investigations along
these lines are ongoing. In Petrov type O or N spaces, the Weyl vector
and Ricci tensor may vanish in different gauges; the dilatational
curvature may not vanish when the original gauge theory action has
$\beta=2\alpha$.

\section*{Acknowledgements}

The author thanks J. Trujillo, J. Hazboun, and C. Torre for illuminating
discussions, an anonymous referee for urging clarification of the
role of torsion, and J. Trujillo for his careful checking of many
of the calculations.

\section*{Appendix A: Structure equations and Bianchi identities}

The curvature $2$-forms are given in terms of the connection by of
the Cartan equations for the conformal group:
\begin{eqnarray}
\boldsymbol{\Omega}_{\; b}^{a} & = & \mathbf{d}\boldsymbol{\omega}_{\; b}^{a}-\boldsymbol{\omega}_{\; b}^{c}\wedge\boldsymbol{\omega}_{\; c}^{a}-2\Delta_{db}^{ac}\mathbf{f}_{c}\wedge\mathbf{e}^{d}\label{Aux spin connection}\\
\mathbf{T}^{a} & = & \mathbf{d}\mathbf{e}^{a}-\mathbf{e}^{b}\wedge\boldsymbol{\omega}_{\; b}^{a}-\boldsymbol{\omega}\wedge\mathbf{e}^{a}\label{Aux solder form}\\
\mathbf{S}_{a} & = & \mathbf{d}\mathbf{f}_{a}-\boldsymbol{\omega}_{\; a}^{b}\wedge\mathbf{f}_{b}+\boldsymbol{\omega}\wedge\mathbf{f}_{a}\label{Aux special conformal}\\
\boldsymbol{\Omega} & = & \mathbf{d}\boldsymbol{\omega}-\mathbf{e}^{a}\wedge\mathbf{f}_{a}\label{Aux dilatation}
\end{eqnarray}
where the principal bundle structure allows each curvature to be expanded
quadratically in the solder forms, e.g.,
\[
\boldsymbol{\Omega}_{\; b}^{a}=\frac{1}{2}\Omega_{\; bcd}^{a}\mathbf{e}^{c}\wedge\mathbf{e}^{d}
\]
Each of the 15 Cartan equations has an integrability condition arising
from the Poincaré lemma, $\mathbf{d}^{2}\equiv0$. For example, the
exterior derivative of the torsion is

\[
\mathbf{d}\mathbf{T}^{a}=\mathbf{d}^{2}\mathbf{e}^{a}-\mathbf{d}\mathbf{e}^{b}\wedge\boldsymbol{\omega}_{\; b}^{a}+\mathbf{e}^{b}\wedge\mathbf{d}\boldsymbol{\omega}_{\; b}^{a}-\mathbf{d}\boldsymbol{\omega}\wedge\mathbf{e}^{a}+\boldsymbol{\omega}\wedge\mathbf{d}\mathbf{e}^{a}
\]
Using the Poincaré lemma to set $\mathbf{d}^{2}\mathbf{e}^{a}\equiv0$,
and substituting from eqs.(\ref{Aux spin connection}), (\ref{Aux solder form})
and (\ref{Aux dilatation}), all terms except those linear in curvatures
cancel, leaving
\begin{eqnarray*}
\mathbf{d}\mathbf{T}^{a} & = & -\mathbf{T}^{b}\wedge\boldsymbol{\omega}_{\; b}^{a}+\boldsymbol{\omega}\wedge\mathbf{T}^{a}+\mathbf{e}^{b}\wedge\boldsymbol{\Omega}_{\; b}^{a}-\boldsymbol{\Omega}\wedge\mathbf{e}^{a}
\end{eqnarray*}
Carrying out similar calculations for the remaining Cartan equations,
eqs.(\ref{Aux spin connection}), (\ref{Aux special conformal}) and
(\ref{Aux dilatation}), we find

\begin{eqnarray}
0 & = & \mathbf{D}\boldsymbol{\Omega}_{\; b}^{a}+2\Delta_{db}^{ac}\left(\boldsymbol{\Omega}_{c}\wedge\mathbf{e}^{d}-\mathbf{f}_{c}\wedge\boldsymbol{\Omega}^{d}\right)\label{Bianchi for rotations}\\
0 & = & \mathbf{D}\mathbf{T}^{a}-\mathbf{e}^{b}\wedge\boldsymbol{\Omega}_{\; b}^{a}+\boldsymbol{\Omega}\wedge\mathbf{e}^{a}\label{Bianchi for torsion}\\
0 & = & \mathbf{D}\mathbf{S}_{a}+\boldsymbol{\Omega}_{\; a}^{b}\wedge\mathbf{f}_{b}-\mathbf{f}_{a}\wedge\boldsymbol{\Omega}\label{Bianchi for co-torsion}\\
0 & = & \mathbf{D}\boldsymbol{\Omega}+\mathbf{T}^{a}\wedge\mathbf{f}_{a}-\mathbf{e}^{a}\wedge\mathbf{S}_{a}\label{Bianchi for dilatations}
\end{eqnarray}
where $\mathbf{D}$ is the Weyl covariant derivative,
\begin{eqnarray*}
\mathbf{D}\boldsymbol{\Omega}_{\; b}^{a} & = & \mathbf{d}\boldsymbol{\Omega}_{\; b}^{a}+\boldsymbol{\Omega}_{\; b}^{c}\wedge\boldsymbol{\omega}_{\; c}^{a}-\boldsymbol{\Omega}_{\; c}^{a}\wedge\boldsymbol{\omega}_{\; b}^{c}\\
\mathbf{D}\mathbf{T}^{a} & = & \mathbf{d}\mathbf{T}^{a}+\mathbf{T}^{b}\wedge\boldsymbol{\omega}_{\; b}^{a}-\boldsymbol{\omega}\wedge\mathbf{T}^{a}\\
\mathbf{D}\mathbf{S}_{a} & = & \mathbf{d}\mathbf{S}_{a}-\boldsymbol{\omega}_{\; a}^{b}\wedge\mathbf{S}_{b}+\mathbf{S}_{a}\wedge\boldsymbol{\omega}\\
\mathbf{D}\boldsymbol{\Omega} & = & \mathbf{d}\boldsymbol{\Omega}
\end{eqnarray*}

When one of the curvature $2$-forms is zero, the Bianchi identities
give algebraic relations. Thus, with vanishing torsion, $\mathbf{T}^{a}=0$,
eq.(\ref{Bianchi for torsion}) becomes
\begin{eqnarray*}
0 & = & -\mathbf{e}^{b}\wedge\boldsymbol{\Omega}_{\; b}^{a}+\boldsymbol{\Omega}\wedge\mathbf{e}^{a}\\
 & = & -\frac{1}{2}\Omega_{\; bcd}^{a}\mathbf{e}^{b}\wedge\mathbf{e}^{c}\wedge\mathbf{e}^{d}+\frac{1}{2}\Omega_{cd}\mathbf{e}^{a}\wedge\mathbf{e}^{c}\wedge\mathbf{e}^{d}\\
 & = & \frac{1}{2}\left(-\Omega_{\; bcd}^{a}+\delta_{b}^{a}\Omega_{cd}\right)\mathbf{e}^{b}\wedge\mathbf{e}^{c}\wedge\mathbf{e}^{d}
\end{eqnarray*}
so that
\[
\Omega_{\;\left[bcd\right]}^{a}=\delta_{[b}^{a}\Omega_{cd]}
\]

\section*{Appendix B: Homogeneous Weyl invariance of the action}

In the linear $SO\left(4,2\right)$ representation, an infinitesmial
conformal transformation takes the form
\[
g_{\; B}^{A}=\delta_{\; B}^{A}+\Lambda_{\; B}^{A}
\]
where $A,B=0,1,\ldots5$. With $a,b=0,1,2,3$, we let $\Lambda_{\; b}^{a}$
be an infinitesimal local Lorentz transformation, $\Lambda^{a}\equiv\Lambda_{\;4}^{a}$
a local translation, $\Lambda_{a}\equiv\Lambda_{\; a}^{4}$ a local
special conformal transformation, and $\Lambda\equiv\Lambda_{\;4}^{4}$
a local dilatation. Antisymmetry of the generators allows us to write
the remaining $\Lambda_{\; B}^{A}$ in terms of these. Then the infinitesimal
gauge transformations of the conformal connection forms are given
by
\begin{eqnarray*}
\delta\boldsymbol{\omega}_{\; b}^{a} & = & \left(\Lambda_{\; c}^{a}\boldsymbol{\omega}_{\; b}^{c}-\boldsymbol{\omega}_{\; c}^{a}\Lambda_{\; b}^{c}\right)+\left(\Lambda^{a}\mathbf{f}_{b}-\mathbf{e}^{a}\Lambda_{b}\right)\\
 &  & +\eta^{ac}\eta_{bd}\left(\Lambda_{c}\mathbf{e}^{d}-\mathbf{f}_{c}\Lambda^{d}\right)-\mathbf{d}\Lambda_{\; b}^{a}\\
\delta\mathbf{e}^{a} & = & \Lambda_{\; c}^{a}\mathbf{e}^{c}+\Lambda^{a}\boldsymbol{\omega}-\boldsymbol{\omega}_{\; c}^{a}\Lambda^{c}-\mathbf{e}^{a}\Lambda-\mathbf{d}\Lambda^{a}\\
\delta\mathbf{f}_{b} & = & \Lambda_{c}\boldsymbol{\omega}_{\; b}^{c}+\Lambda\mathbf{f}_{b}-\mathbf{f}_{c}\Lambda_{\; b}^{c}-\boldsymbol{\omega}\Lambda_{b}-\mathbf{d}\Lambda_{b}\\
\delta\boldsymbol{\omega} & = & \Lambda_{c}\mathbf{e}^{c}-\mathbf{f}_{c}\Lambda^{c}-\mathbf{d}\Lambda
\end{eqnarray*}
The auxiliary gauging breaks the translational symmetry. Without the
translations these reduce to
\begin{eqnarray*}
\delta\boldsymbol{\omega}_{\; b}^{a} & = & \Lambda_{\; c}^{a}\boldsymbol{\omega}_{\; b}^{c}-\boldsymbol{\omega}_{\; c}^{a}\Lambda_{\; b}^{c}-\mathbf{e}^{a}\Lambda_{b}+\eta^{ac}\eta_{bd}\Lambda_{c}\mathbf{e}^{c}-\mathbf{d}\Lambda_{\; b}^{a}\\
\delta\mathbf{e}^{a} & = & \Lambda_{\; c}^{a}\mathbf{e}^{c}-\mathbf{e}^{a}\Lambda\\
\delta\mathbf{f}_{b} & = & \Lambda_{c}\boldsymbol{\omega}_{\; b}^{c}+\Lambda\mathbf{f}_{b}-\mathbf{f}_{c}\Lambda_{\; b}^{c}-\boldsymbol{\omega}\Lambda_{b}-\mathbf{d}\Lambda_{b}\\
\delta\boldsymbol{\omega} & = & \Lambda_{c}\mathbf{e}^{c}-\mathbf{d}\Lambda
\end{eqnarray*}
showing that the solder form has become tensorial.

For the curvatures, the $\mathcal{IW}$ transformations are similar,
\begin{eqnarray*}
\delta\boldsymbol{\Omega}_{\; b}^{a} & = & \Lambda_{\; c}^{a}\boldsymbol{\Omega}_{\; b}^{c}-\boldsymbol{\Omega}_{\; c}^{a}\Lambda_{\; b}^{c}-\mathbf{T}^{a}\Lambda_{b}+\eta^{ac}\eta_{bd}\Lambda_{c}\mathbf{T}^{d}\\
\delta\mathbf{T}^{a} & = & \Lambda_{\; c}^{a}\mathbf{T}^{c}-\mathbf{T}^{a}\Lambda\\
\delta\mathbf{S}_{b} & = & \Lambda_{c}\boldsymbol{\Omega}_{\; b}^{c}+\Lambda\mathbf{S}_{b}-\mathbf{S}_{c}\Lambda_{\; b}^{c}-\boldsymbol{\Omega}\Lambda_{b}\\
\delta\boldsymbol{\Omega} & = & \Lambda_{c}\mathbf{T}^{c}
\end{eqnarray*}
Notice that if we suppress special conformal transformations, $\Lambda_{a}=0$,
both the Lorentz curvature and dilatational curvature become separate
tensors under the remaining $\mathcal{W}$ transformations so the
action eq.(\ref{eq: W invariant action}) is manifestly $\mathcal{W}$-invariant.
The translational symmetry has been replaced by general coordinate
invariance of the curved manifold \cite{Kibble}.

For a full $\mathcal{IW}$ transformation of the action we easily
show that
\begin{eqnarray*}
\delta S_{auxiliary}^{\mathcal{W}} & = & -4\alpha\int\Lambda_{b}\mathbf{T}^{a}\wedge{}^{*}\boldsymbol{\Omega}_{\; a}^{b}+2\beta\int\Lambda_{c}\mathbf{T}^{c}\wedge{}^{*}\boldsymbol{\Omega}
\end{eqnarray*}
Therefore, $S_{auxiliary}^{\mathcal{W}}$ is invariant if we perform
no special conformal transformations, $\Lambda_{b}=0$, or if the
torsion vanishes, $\mathbf{T}^{a}=0$.

\section*{Appendix C: The basis structure equation and the connection}

Various relations between the solder form, metric, and connection
are readily established from the basis structure equation, eq.(\ref{Aux solder form}),
\[
\mathbf{d}\mathbf{e}^{a}=\mathbf{e}^{b}\wedge\boldsymbol{\omega}_{\; b}^{a}+\boldsymbol{\omega}\wedge\mathbf{e}^{a}+\mathbf{T}^{a}
\]
Eq.(\ref{Aux solder form}) gives the antisymmetric part of the partial
derivative of the solder form, so we must have

\[
\partial_{\nu}e_{\mu}^{\quad a}+e_{\mu}^{\quad b}\omega_{\; b\nu}^{a}-W_{\nu}e_{\mu}^{\quad a}+\frac{1}{2}T_{\;\mu\nu}^{a}=\Sigma_{\;\mu\nu}^{a}
\]
where $\Sigma_{\;\mu\nu}^{a}$ is symmetric, $\Sigma_{\;\mu\nu}^{a}=\Sigma_{\;\nu\mu}^{a}$.
Permuting indices and combining to solve for $\Sigma_{\;\mu\nu}^{a}$
in the usual way leads to 
\begin{eqnarray*}
\Sigma_{\mu\alpha\nu} & = & \frac{1}{2}\left(\partial_{\nu}g_{\alpha\mu}+\partial_{\alpha}g_{\mu\nu}-\partial_{\mu}g_{\nu\alpha}\right)-\left(g_{\alpha\mu}W_{\nu}+g_{\mu\nu}W_{\alpha}-g_{\nu\alpha}W_{\mu}\right)+\frac{1}{2}\left(T_{\alpha\mu\nu}+T_{\nu\mu\alpha}\right)
\end{eqnarray*}
and substituting back into the derivative of the solder form, we have
\[
D_{\nu}e_{\mu}^{\quad a}\equiv\partial_{\nu}e_{\mu}^{\quad a}+e_{\mu}^{\quad b}\omega_{\; b\nu}^{a}-e_{\alpha}^{\quad a}\tilde{\Gamma}_{\;\mu\nu}^{\alpha}-e_{\mu}^{\quad a}W_{\nu}=0
\]
where we define the connection (for a Weyl geometry with torsion)
as
\[
\tilde{\Gamma}_{\;\mu\nu}^{\beta}\equiv\Gamma_{\;\mu\nu}^{\beta}-\left(\delta_{\mu}^{\beta}W_{\nu}+\delta_{\nu}^{\beta}W_{\mu}-g^{\alpha\beta}g_{\nu\mu}W_{\alpha}\right)+\frac{1}{2}\left(T_{\mu\;\nu}^{\;\beta}+T_{\nu\;\mu}^{\;\beta}-T_{\;\mu\nu}^{\beta}\right)
\]
As expected,
\[
\tilde{\Gamma}_{\;\mu\nu}^{\beta}-\tilde{\Gamma}_{\;\nu\mu}^{\beta}=-T_{\;\mu\nu}^{\beta}
\]
Contracting $D_{\nu}e_{\mu}^{\quad a}$ with a second solder form
and symmetrizing, we have metric compatibility,
\[
D_{\nu}g_{\alpha\mu}=\partial_{\nu}g_{\alpha\mu}-g_{\alpha\beta}\tilde{\Gamma}_{\;\mu\nu}^{\beta}-g_{\mu\beta}\tilde{\Gamma}_{\;\alpha\nu}^{\beta}=0
\]
and standard manipulations (e.g., \cite{Weinberg}) show that
\[
\partial_{\alpha}\left(\sqrt{-g}\right)=\sqrt{-g}\tilde{\Gamma}_{\;\mu\alpha}^{\mu}
\]

\end{document}